\documentclass[aps,prb,twocolumn,superscriptaddress,address,amsmath,amssymb,showpacs]{revtex4}

\usepackage{graphicx}

\begin{document}


\title{Magnetically robust topological edge states and flat bands}

\author{Tomi Paananen and Thomas Dahm}

\affiliation{Universit\"at Bielefeld, Fakult\"at f\"ur Physik, 
Postfach 100131, D-33501 Bielefeld, Germany}

\date{\today}

\begin{abstract}
We study thin strips of three dimensional topological insulators
in the presence of a spin-splitting Zeeman field. We show that
under certain conditions the topological edge states at the
sides of a strip remain robust against a time-reversal
symmetry breaking Zeeman field. For a particle-hole symmetric
system with Zeeman field perpendicular to the strip we strictly 
proof that the dispersion and the spin-orbital structure of the edge states 
remains unchanged.
When the Zeeman field lies parallel to the strip, the Dirac
spectrum becomes flat, but remains intact. Above a critical value
of the Zeeman field a topological flat band appears at the edge.
We present numerical calculations for a lattice model
of Bi$_2$Se$_3$. These calculations show that even though
particle-hole symmetry is not strictly fulfilled in this
system, these special features are still present.
The flat band is tunable by the Zeeman field and can
be realistically achieved in Bi$_2$Se$_3$-ferromagnet
heterosystems at room temperature.
\end{abstract}

\pacs{73.20.At, 75.70.-i, 73.43.-f, 03.65.Vf}


\maketitle

\section{Introduction}

Symmetries play an important role in condensed matter physics.
Recently, a peculiar state of matter, the topological insulator,
has been first suggested theoretically \cite{Bernevig,Fu} and 
subsequently observed experimentally. 
\cite{Koenig,Hsieh1,Chen,Xia,Hsieh2,Kuroda} 
In these materials time-reversal symmetry
guarantees protection of conducting surface or edge states against
backscattering, which is a requirement for dissipationless
spin transport.
Usually, when a time-reversal symmetry breaking perturbation is
applied, like a magnetic field for example, the edge states
are not protected anymore and may acquire a gap. \cite{Chu:PRB11} 
Protected topological surface states in time-reversal symmetry
broken systems have recently been suggested in superfluid
$^3$He-B \cite{Machida} and topological superconductors. \cite{PALee}
Here, we will show that under certain circumstances even
edge states of a topological insulator may become
robust under a time-reversal symmetry breaking perturbation.
In particular, a robust flat band may appear.


Recently, several materials, like Bi$_2$Se$_3$, Bi$_2$Te$_3$,
Sb$_2$Te$_3$, or PbBi$_2$Te$_4$ have been experimentally identified
as three dimensional topological insulators (3DTI). 
\cite{Chen,Xia,Hsieh2,Kuroda}
The topological protection of the surface states and their spin-momentum
coupled nature makes these
systems particularly interesting for spintronics. Several
devices have been proposed theoretically. 
\cite{Garate,Yokoyama,Yu:Science10,Hosur,BlackSchaffer} In particular, the
combination with ferromagnetic materials or the application
of magnetic fields can lead to interesting effects.

In the present work we study thin strips of 3DTI materials
in the presence of a time-reversal
symmetry breaking Zeeman field. Zeeman fields can be
introduced into topological insulators either by doping
with ferromagnetic dopants \cite{Yu:Science10} or by proximity to a
ferromagnetic material (see Fig. \ref{Fig01}). We show that in this case
certain additional spin-orbital
symmetries are present, which are respected by the
Zeeman field. These symmetries can make the
edge states robust even against a Zeeman field. We 
will also show that at larger Zeeman fields a phase
transition appears into a state with a robust flat band.
Flat bands in solids are particularly interesting objects,
because the group velocity vanishes, allowing for
immobile wave packets, localized states, and a
giant effective mass. \cite{Nishino} The special feature
of the flat band we find here is that it is tunable by
the Zeeman field and can be realistically achieved in
Bi$_2$Se$_3$ thin films.

\section{Model}

\begin{figure}[t]
\includegraphics[width=0.98 \columnwidth]{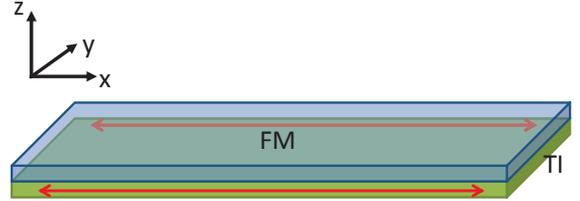} 
\caption{\label{Fig01} (Color online)
Geometry of the thin strip considered here: the strip (green, TI) has a 
finite extension in $y$-direction and is confined
in $z$-direction, such that the edge states (red) become dominant
at the Fermi level.
An in-plane or out-of-plane Zeeman field can be provided by
a ferromagnetic layer (blue, FM) via the proximity effect, for example.
}
\end{figure}

We start from the lattice regularized version of an effective 
two-orbital model for
three dimensional topological insulators suggested by
Li et al \cite{Li:NPhys10} in the presence of a Zeeman field.
This is one of the two models that are currently being discussed for
3DTI materials. \cite{TKLee} These two models differ only in the way 
how the spin-orbit interaction couples in $z$-direction. For a 
thin strip confined in $z$-direction as considered here, both 
models yield the same effective two-dimensional Hamiltonian.
We are interested in strips sufficiently thin such that the top and bottom
surface states are gapped out and the edge states become dominant at the
Fermi level. (For Bi$_2$Se$_3$ this would mean a film
thickness of less than 6~nm. \cite{Zhang:NPhys10}) In this case
the confinement in $z$-direction leads to a generic effective two-dimensional
Hamiltonian of the form:
\begin{equation} \label{eq:hamiltonian}
	H(\mathbf k)=\epsilon_0(\mathbf k) \mathbb{I}_{4 \times 4} +
\sum_{i=0}^2 m_i(\mathbf k) \Gamma^i + \sum_{\alpha \in \{x,y,z \}} V_\alpha
\; \mathbb{I}_{2 \times 2} \otimes \sigma_\alpha
\end{equation}
with $\epsilon_0(\mathbf k) = C + 2 D ( 1- \cos k_x )+ 2 D ( 1- \cos k_y )$,
$m_0(\mathbf k) = M_{2D} - 2 B ( 1- \cos k_x ) - 2 B ( 1- \cos k_y )$,
$m_1(\mathbf k) = 2 A \sin k_x$, and $m_2(\mathbf k) = 2 A \sin k_y$.
Here, the Dirac $\Gamma$ matrices are represented by
$\Gamma^{0,1,2}=(\tau_z \otimes \mathbb{I}_{2 \times 2}, \tau_x \otimes \sigma_x, 
\tau_x \otimes \sigma_y)$ in the basis of bonding and antibonding
$p_z$ orbital states. The Pauli matrices in orbital space
are denoted by $\tau_i$ and the ones in spin space by $\sigma_i$.
The components of the Zeeman field in $x$, $y$, and $z$-direction
are denoted by $V_{x,y,z}$, respectively. The parameters $A$, $B$, $C$,
$D$, and $M$ have been derived from bandstructure calculations
for Bi$_2$Se$_3$ in Ref.~\onlinecite{Zhang:NPhys09}. Our parameter
$M_{2D}=M-B_1 (\pi/L_z)^2$ is an effective two dimensional parameter,
which accounts for the finite film thickness $L_z$. 

First, we will study a particle-hole symmetric system with $D=0$.
In Bi$_2$Se$_3$ the parameter $D$ is about a factor of three smaller than $B$
and its modest influence will be considered in Section \ref{sec3}. 
Without loss of 
generality we can set $C=0$. The Hamiltonian (\ref{eq:hamiltonian}) then
possesses several spin-orbital symmetries in the absence of a Zeeman
field some of which remain valid for certain directions of the Zeeman field.
Specifically, let us consider the following symmetry operators:
$\Theta_0=\tau_z \otimes \sigma_x$, $\Theta_1=\tau_x \otimes \sigma_x$, 
$\Theta_2=\tau_y \otimes \sigma_y$, $\Theta_3=\tau_z \otimes \sigma_z$,
and parity $P=\tau_z \otimes \mathbb{I}_{2 \times 2}$. These symmetry
operators share that they act on both orbital and spin degrees of 
freedom at the same time. 
In the absence of a Zeeman field the Hamiltonian (\ref{eq:hamiltonian})
fulfils the following symmetry relations:
$\Theta_0 H(k_x,k_y) \Theta_0^{-1} = H(-k_x,k_y)$,
$\Theta_1 H(k_x,k_y) \Theta_1^{-1} = -H(-k_x,k_y)$,
$\Theta_2 H(k_x,k_y) \Theta_2^{-1} = -H(-k_x,k_y)$,
$\Theta_3 H(k_x,k_y) \Theta_3^{-1} = H(k_x,k_y)$, and
$P H(k_x,k_y) P^{-1} = H(-k_x,-k_y)$.
In the presence of a Zeeman field in $z$-direction the symmetries
$\Theta_1$, $\Theta_2$, $\Theta_3$, and $P$ are still respected, while $\Theta_0$
is broken. In contrast, in the presence of a Zeeman field 
in $x$-direction the symmetries $\Theta_0$, $\Theta_2$, and $P$ are still 
respected, while $\Theta_1$ and $\Theta_3$ are broken.
All of these symmetry operations respect $\Theta_i^2=P^2=\mathbb{I}_{4 \times 4}$,
thus their eigenvalues are $\pm 1$.

Let us consider now a strip with finite extension in $y$-direction
with $y \in [0,L]$, but periodic boundary conditions in $x$-direction,
as shown in Fig.~\ref{Fig01}.
In the absence of a Zeeman field the system possesses four topological 
edge states at the two edges $y=0$ and $y=L$, two on each side.
These edge states can be obtained analytically in small $k_y$ expansion
following the method of Ref.~\onlinecite{BZhou}. If $L$ is sufficiently
larger than the localization length of the edgestates, we may consider
the edges separately letting $L \rightarrow \infty$.
The edge state wavefunctions are then linear superpositions of 
exponentially decaying functions of the form $e^{\eta y}$ with the 
boundary conditions $\psi(y=0)=0$ and $\psi(y \rightarrow \infty)=0$. 

\subsection{Zeeman field in $z$-direction: robust edge states}

We will first proof
that the dispersion of the edge states remains unaffected by a
Zeeman field in $z$-direction. Consider the Hamiltonian
\begin{equation} \label{eq:h0}
	H_0(\mathbf k)=m_0(\mathbf k) \Gamma^0 + m_2(\mathbf k) \Gamma^2 
+ V_z\; \mathbb{I}_{2 \times 2} \otimes \sigma_z
\end{equation}
This is the Hamiltonian (\ref{eq:hamiltonian}) with $C=D=0$
except for the $\Gamma^1$
term. Note, that this Hamiltonian anticommutes with the symmetry
operation $\Theta_1$, i.e. $\left[ H_0, \Theta_1 \right]_+=0$. 
Now assume that $\left| \Psi_{k_x} \right\rangle$ is a zero-energy eigenstate
of $H_0$. It is easy to show by direct calculation that such a
zero-energy edge state exists, if $M_{2D} - 2 B ( 1- \cos k_x ) +V_z > 0$. 
A second one exists, if $M_{2D} - 2 B ( 1- \cos k_x ) -V_z > 0$.
Physically, this means that an edge state exist for
each of the two spin-split bulk bands whenever there
is a band inversion. \cite{Bernevig}
Equivalently, one can show that the Hamiltonian (\ref{eq:h0})
possesses two flat bands protected by the symmetry $\Theta_1$,
one for each of the two bulk bands using the topological
invariant in Ref.~\onlinecite{Matsuura}.
Due to the anticommutation, $\left| \Psi_{k_x} \right\rangle$
either is an eigenstate of $\Theta_1$ or can be constructed as such.
Now it turns out that $\Theta_1=\Gamma^1$. Therefore 
$\left| \Psi_{k_x} \right\rangle$ will also be an eigenstate of
the Hamiltonian (\ref{eq:hamiltonian}) (with $C=D=0$) and its
energy dispersion will thus be given by the prefactor of the
$\Gamma^1$ term times the $\Theta_1$~eigenvalue, 
i.e. $E(k_x)=\pm 2 A \sin k_x$. This result is
apparently independent of $V_z$ (and also of $M_{2D}$ and $B$,
as long as the condition for the existence of the edge state
is fulfilled). As the Hamiltonian also commutes with $\Theta_3$
we can deduce that the four edge states have to be proportional
to $(1,0,0,1)$, $(1,0,0,-1)$, $(0,1,1,0)$, or $(0,1,-1,0)$,
i.e. also the spin-orbital structure of the edge states
remains unaffected by $V_z$. It is only the spatial structure
that changes. Note, that this spin-orbital structure has
the property that the expectation value of the total spin-$z$
component vanishes, because the spin in the two orbital subbands is just
opposite. This is in contrast to the quantum spin Hall system 
HgTe, where the edge states consist of two orbital
subbands having the same spin. \cite{BZhou,TKLee} In the
present case, which is relevant for the edge states in
a thin Bi$_2$Se$_3$ strip, this special feature explains
why the edge states are robust against a Zeeman field
in $z$-direction. The robustness of the edge states can
equally be understood from the existence of a flat band
in the corresponding Hamiltonian (\ref{eq:h0}).

The parity operator $P$ relates a state at one
edge to a state with same energy and opposite $k_x$ on
the other edge. From this we can conclude that the states
$(1,0,0,1)$ and $(1,0,0,-1)$ are localized on different
edges and have the same spatial structure and the
states $(0,1,1,0)$ and $(0,1,-1,0)$, too. 
For $V_z=0$
the symmetry $\Theta_0\Theta_2$ shows that
$(1,0,0,1)$ and $(0,1,-1,0)$ are on the same edge
with same momentum, but opposite energy. The same is true
for $(1,0,0,-1)$ and $(0,1,1,0)$. These statements also
hold for finite $V_z$ even though $\Theta_0$ is broken.

\begin{figure}[t]
\includegraphics[width=0.98 \columnwidth]{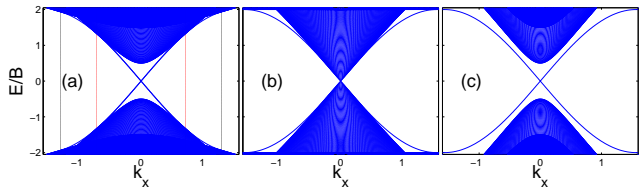} 
\caption{\label{Fig02} (Color online)
Numerical dispersions of bulk and edge states for 
$V_z/M_{2D}=0.5$, $1.0$, and $1.5$, respectively.
When $V_z/M_{2D}$ reaches $1$ two of the four edge states disappear.
We have used $B=A=M_{2D}=1$, and $C=D=0$. The vertical lines
in (a) show the momenta $k_x$, where 
$M_{2D} - 2 B ( 1- \cos k_x ) \pm V_z = 0$.
}
\end{figure}

In order to confirm our analytical results we have diagonalized
the Hamiltonian (\ref{eq:hamiltonian}) with $C=D=0$
and $M_{2D}=B=A=1$
on a finite lattice of size $500 \times 200$ numerically with periodical boundary conditions in $x$-direction.
The numerical results, shown in Fig.~\ref{Fig02} 
confirm the analytical proof.
For $V_z/M_{2D}<1$ we find four edge states, two on each side.
When $V_z/M_{2D}$ approaches 1, the localization length of two of the 
edge states approaches the width of the system, similar to
what has been found in Ref.~\onlinecite{Wada}.
At $V_z/M_{2D}=1$ the bulk bands touch the Dirac point (see Fig.~\ref{Fig02}(b)).
When $V_z/M_{2D}$ becomes larger than 1, these two edge states
disappear. The other two edge states
survive, though. The system thus makes a quantum phase
transition into a quantum anomalous Hall state.
Throughout this transition the dispersion and the spin-orbital
structure of the edge states remains unchanged, while the 
bulk spectrum changes, of course.

\subsection{Zeeman field in $x$-direction: appearence of a flat band}

\begin{figure}[t]
 \includegraphics[width=0.98 \columnwidth]{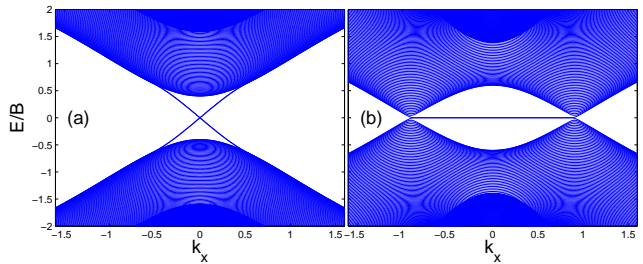}
\caption{\label{Fig03} (Color online)
Numerical dispersions of bulk and edge states for a Zeeman field
in $x$-direction with $V_x/M_{2D}=0.6$, and $1.6$, respectively.
When $V_x/M_{2D}$ exceeds $1$ two of the four edge states disappear and a flat band appears.  
Parameters are the same as in figure \ref{Fig02}.
}
\end{figure}

Let us next consider a Zeeman field in $x$-direction.
In this case the dispersion of the edge states changes,
because $\mathbb{I}_{2 \times 2} \otimes \sigma_x$ does not
anticommute with $\Theta_1$ anymore.
The symmetries $\Theta_0$ and $\Theta_2$ are still 
respected. Let $\left| \Psi_{k_x} \right\rangle$ be an
edge state of $H$ with energy $E$. Then,
$\Theta_0 \Theta_2 \left| \Psi_{k_x} \right\rangle$
is an edge state with energy $-E$ which is localized
at the same edge. If we have just one edge state on each side,
$\left| \Psi_{k_x} \right\rangle= \pm \Theta_0 \Theta_2 \left|
\Psi_{k_x}\right\rangle$
must hold and its energy thus has to be strictly zero,
i.e. a flat band appears. As shown in Fig.~\ref{Fig03},
this happens for $V_x>\sqrt{M(k_x)^2+4A^2\sin^2(k_x)}$,
where $M(k_x)=M_{2D} - 2 B ( 1- \cos k_x )$.
This is a topological flat band as has been
discussed recently in graphene \cite{Ryu} and topological
superconductors with gap nodes. \cite{Schnyder} Such a flat band
appears when the bulk gap closes, \cite{Volovik} as is the case here, too.
Its existence is guaranteed by a topological invariant due
to the chiral symmetry $\Theta_0 \Theta_2$. \cite{Matsuura,Schnyder}
In contrast to other flat band systems, however, in the present case 
the flat band is generated by a magnetic exchange field and thus is 
tunable experimentally. We 
will show below that it can be observed realistically in Bi$_2$Se$_3$.

In Fig.~\ref{Fig03} we show the dispersions of bulk and edge states above
and below $V_x/M_{2D}=1$. For $V_x/M_{2D}<1$ we have four edge states. When $V_x/M_{2D}>1$
two of these states disappear and a pair of bulk Dirac points appears.
These points are connected by a flat band of two edge states, one on each side
of the strip.

When the Zeeman field is rotated within the $x$-$y$-plane the
flat band becomes smaller and finally shrinks to a single
point, when the Zeeman field points in $y$-direction.

\section{\label{sec3}Realistic parameters for B\lowercase{i}$_2$S\lowercase{e}$_3$}

So far we have discussed the particle-hole symmetric case $C=D=0$ in
Eq.~(\ref{eq:hamiltonian}), in which the two phenomena of stable
dispersion and flat bands become particularly clear.
We will now study parameters, which are realistic for Bi$_2$Se$_3$,
in order to show that these phenomena remain and how they are modified.
According to Zhang et al. \cite{Zhang:NPhys09} the parameters are given
by $B=3.3\,\text{eV}$, $M_{2D}= 0.17\,\text{eV}$, $A= 0.49\,\text{eV}$, $D= 1.14\,\text{eV}$, and $C=-0.0068\,\text{eV}$,
i.e. particle-hole symmetry is somewhat broken. Here, we considered
a film thickness of 3~nm and took the lattice constant $a=4.14$~\AA.
The parameter $M_{2D}$ is the size
of the bulk gap in the absence of a Zeeman field. The parameter
$t=D/B=0.35$ is a measure of the particle-hole asymmetry. 
We note that the particle-hole broken Hamiltonian is adiabatically connected
to the particle-hole symmetric one, if $|t|<1$. This means that the topological
phase is the same in both cases.
For $|t|>1$ the system would become topologically trivial.

Following the method of Zhou et al. \cite{BZhou} we can determine the
dispersion and structure of the edge states analytically, when a
Zeeman field in $z$-direction is applied. For the dispersions we find
$E(k_x)=t (M_{2D} \pm V_z ) \pm 2 A \sqrt{1-t^2} \sin k_x$.
The spin-orbital structure of the edge states turns out to be of
the form $(u,0,0,v)$, $(u,0,0,-v)$, $(0,v,u,0)$, and $(0,-v,u,0)$,
where $u=\sqrt{1-t^2}$ and $v=\sqrt{1+t^2}$. For $t=0$ the particle-hole
symmetric results above are reproduced. Interestingly, the dispersions remain robust
except for a constant energy shift of $\pm t V_z$, which is
reduced by the particle-hole asymmetry. The Zeeman field does
not create a gap in the dispersion of the edge states. Also, the 
spin-orbital structure of the edge states neither depends on
$V_z$ nor on $k_x$. Only the parameter $t$ determines the
spin-orbital structure.

\begin{figure}[t]
\includegraphics[width=0.98 \columnwidth]{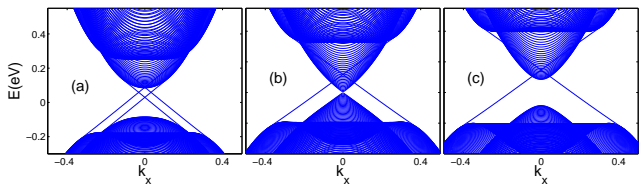}
\caption{\label{Fig04} (Color online)
Numerical dispersions of bulk and edge states for Zeeman field in
$z$-direction with $V_z/M_{2D}=0.5$, $1.0$ and $1.5$, respectively.
When $V_z/M_{2D}$ exceeds $1$, two of the four edge states disappear.  
Here, we have chosen realistic parameters for Bi$_2$Se$_3$, as described in
the text.
}
\end{figure}

In Fig.~\ref{Fig04} we present numerical dispersions for a finite
Bi$_2$Se$_3$ strip in the presence of a Zeeman field in
$z$-direction using the parameters of Zhang et al, which
corroborate our analytical results. Fig.~\ref{Fig04}(a)
shows that $V_z$ shifts the energies of the edge
states, but does not change the shape of the dispersions.
When $V_z/M_{2D}$ exceeds $1$, two of the four edge states
disappear and the system makes a quantum phase
transition into a quantum anomalous Hall state
similar to the one that has recently been found in Bi(111) bilayers.
\cite{HZhang}

\begin{figure}[t]
\begin{tabular}{ll}
 \includegraphics[width=0.49 \columnwidth]{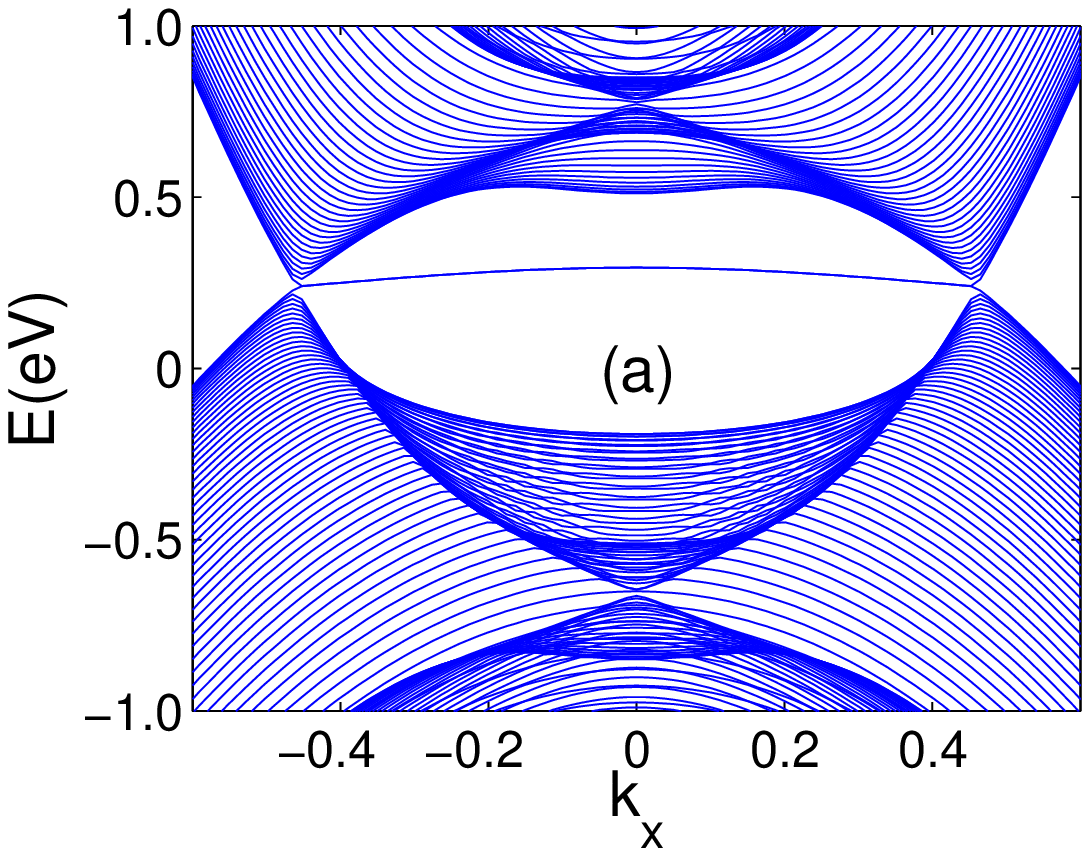}
& \includegraphics[width=0.49 \columnwidth]{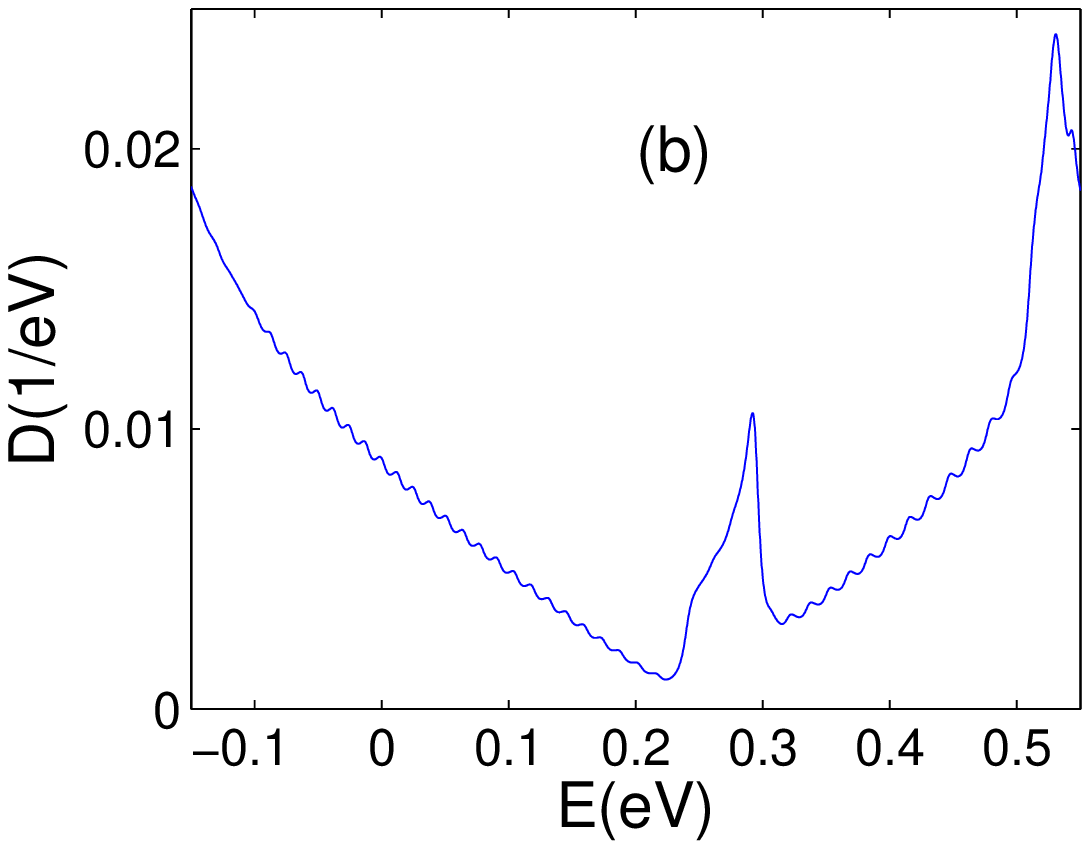}
\end{tabular}
\caption{\label{Fig05} (Color online)
(a) Numerical dispersion of bulk and edge states for a Zeeman field in
$x$-direction with $V_x/M_{2D}=4$.
When $V_x/M_{2D}$ exceeds $1$, two of the four edge states disappear and an
almost flat band appears.  
(b) Density of States of this system.
The flat band appears as a peak around $E=0.28\,\text{eV}$.
Parameters are the same as in figure~\ref{Fig03}.
}
\end{figure}

When the Zeeman field is applied in $x$-direction, the symmetries
$\Theta_0$ and $P$ are still respected, while $\Theta_2$ is broken.
Let $\left| \Psi_{k_x} \right\rangle$ be an
edge state of $H$ with energy $E$. Then,
$\Theta_0 \left| \Psi_{k_x} \right\rangle$
is an edge state with energy $E$ and momentum $-k_x$, 
which is localized at the same edge. If there
exists only one state at each edge we have $E(k_x)=E(-k_x)$,
i.e. the dispersion is symmetric in $k_x$.
The state $P \Theta_0 \left| \Psi_{k_x} \right\rangle$
then is an edge state with same energy and momentum $k_x$ localized
at the opposite edge and the edge state dispersion is twofold
degenerate.

In Fig.~\ref{Fig05}(a) we show the numerical dispersion of bulk and
edge states for a Zeeman field in
$x$-direction with $V_x/M_{2D}=4$. When $V_x/M_{2D}$ exceeds $1$, again two of the
four edge states disappear and an almost flat band appears. 
Due to the broken particle-hole symmetry this band
has a finite dispersion now. For $V_x/M_{2D}=4$ this band is clearly separated
from the bulk bands. It ends at the two Dirac points of the bulk bands,
as in the particle-hole symmetric case. In Fig.~\ref{Fig05}(b) we show
the density of states. Here, the flat band
appears as a peak around $E=0.28$~eV. In order to realize this
situation experimentally one would need Zeeman fields in the range
$0.17\,\text{eV} < V_x < 0.7\,\text{eV}$, which are typical exchange 
splittings in common ferromagnets. The criterion for the
appearance of this flat band remains the same as in the
particle-hole symmetric case above,
i.e. $V_x>\sqrt{M(k_x)^2+4A^2\sin^2(k_x)}$, because its
appearance is related to the closure of the bulk gap, which
is unchanged, when $D>0$. The dispersion of the band decreases
with increasing value of $V_x$.

The robustness of flat bands with respect to disorder has been discussed
recently by Matsuura et al. \cite{Matsuura} In order to check the
robustness, we have added a single impurity at the edge of our system.
The numerical calculation showed that the number of zero energy
states remained unchanged, confirming the robustness of the flat band.
This is expected as the flat bands are protected by a bulk
topological invariant.

\section{Conclusion}

In conclusion, we have shown that spin-orbital symmetries in a topological
insulator can create special features like robust edge states and flat bands.
Specifically, we have studied a thin Bi$_2$Se$_3$ strip in a Zeeman field.
When the Zeeman field is applied perpendicular to the strip we have shown
that the edge states remain robust even under the time-reversal symmetry
breaking Zeeman field. 
When the Zeeman field is applied parallel to the strip along its long
direction, a topological flat band appears above a critical value
of the Zeeman field. In contrast to previous proposals this flat band
is within experimental reach at room temperature and tunable by the 
Zeeman field. The combination of Bi$_2$Se$_3$ thin films with
ferromagnetic materials can thus create highly interesting features
that can be tuned by the polarization of the ferromagnet and
lead to new spintronics devices.

\acknowledgments

We would like to thank A.~P.~Schnyder for valuable discussions.

\end{document}